\documentclass[letterpaper, 10 pt, conference]{ieeeconf}  

\usepackage{graphicx}      
\newtheorem{remark}{\bf Remark}
\newtheorem{theorem}{\bf Theorem}[section]

\newtheorem{proposition}[theorem]{\bf Proposition}
\newcommand{\barr}[2]{\begin{array}{#1}#2\end{array}}
\newcommand{\beq}{\begin{equation}}
\newcommand{\eeq}{\end{equation}}

\newcommand{\R}{\rm{I\kern-2pt R}}

\IEEEoverridecommandlockouts                              

\title{\LARGE \bf
Negotiating Highway Interchange Traffic with a Decentralized Instability-Driven CBF-based Algorithm
}

\author{Mrdjan Jankovic$^{1}$, Shreshta Rajakumar Deshpande$^{2}$, Gopika Ajaykumar$^{2}$
\thanks{$^{1}$Southwest Research Institute,
        Ann Arbor, MI 48108, USA
        {\tt\small mrdjan.jankovic@swri.org}}%
\thanks{$^{2}$Southwest Research Institute, San Antonio, TX 78238, USA
        {\tt\small shreshta.rajakumardeshpande@swri.org, gopika.ajaykumar@swri.org}}%
}

\begin{document}

\maketitle
\thispagestyle{empty}
\pagestyle{empty}

\begin{abstract}
In this paper we consider an interchange lane-swap scenario, a limited stretch of highway with two parallel lanes where most vehicles want to change lanes. We show that a particular decentralized Control Barrier Function based algorithm executes lane swaps efficiently, with minimal speed change, within the specified (short) road segment at high traffic densities (3,500 veh./hour per lane). Our main point is that controller tuning -- the speed of inter-agent instability -- plays a major role in the performance of the vehicle group. This is illustrated by comparing two different tunings of the controller and a third one where the lane swap is enforced by ``virtual guard rails." Like fighter jet dynamic instability improving maneuverability, the inter-agent instability improves agility of a group of vehicles. We emphasize that the controllers considered are decentralized: agents do not know if others want to change lanes or not.   

\end{abstract}

\section{INTRODUCTION}
Highway merging for connected and automated vehicles (CAVs) has been a subject of significant research over the last few decades as reviewed in several survey papers, e.g. \cite{scarinci, rios-torres}. This is motivated in part by disproportionally higher percent of highway accidents happening on on- and off-ramps \cite{mccartt}. There is also a promise of reducing delays, improving traffic flow and fuel efficiency by advanced algorithms and fast communication between connected vehicles.

Among planning-based methods, central control with Model Predictive Control (MPC) and Mixed Integer Program (MIP) has been used \cite{mukai, xu, hayashi}. A sizable fraction of the literature on automated highway merges deal with the one-degree-of freedom (1-DoF) operation where the vehicles are ``on rails" -- that is, steering keeps vehicles in the lane, while the merge order is controlled by acceleration. In most cases the algorithms are decentralized, and often require additional supporting framework, typically the first-in-first-out (FIFO) rule   
\cite{rios-torres_merge, xiao2020decentralized, xu_feasibility}. A 1-DoF merge has been considered with a centralized algorithm without FIFO or other priority structure in \cite{deshpandeMECC}.

\begin{figure}[!b]
\vspace{-3mm}
	\centering
	\includegraphics[width=\columnwidth]{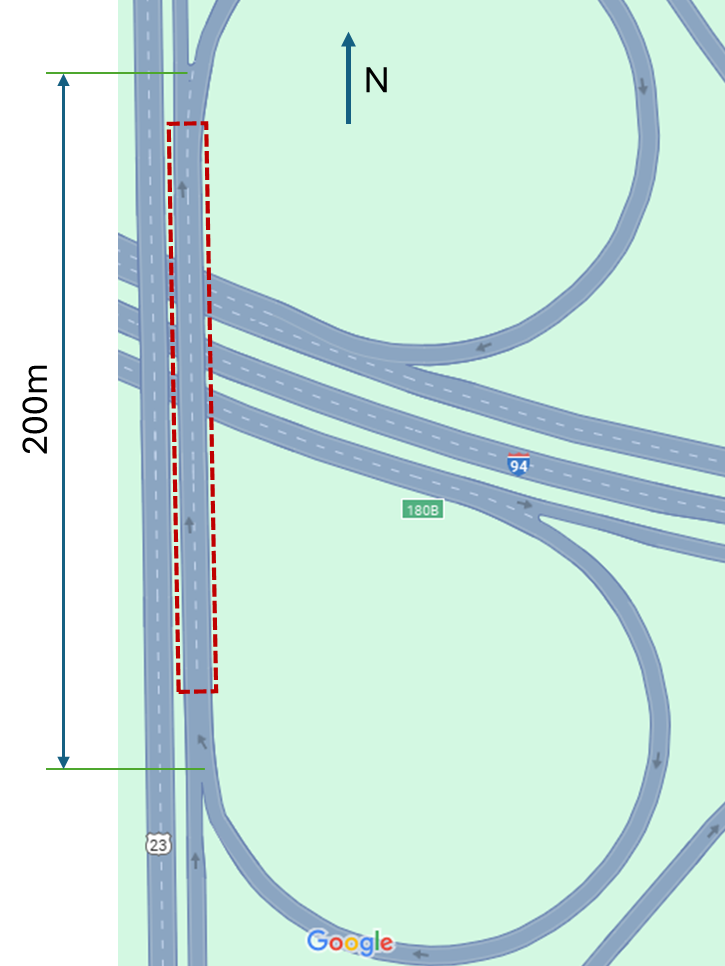}
	\caption{The interchange zone between M-23 and I-94 highways in Michigan near Ann Arbor (source:  Google Maps).}
	\label{fig:interchange}
\end{figure}

In this paper we consider a two degree-of-freedom (2-DoF) lane swaps at dedicated interchanges popular in SE Michigan. One example in the Ann Arbor area is shown in Fig. \ref{fig:interchange}. Here, vehicles leaving M-23 North to merge onto the I-94 West have to use the dedicated interchange, while the I-94 East vehicles that want to go onto M-23 North enter the right lane in the same interchange zone (red-dash rectangle). Most of the vehicles in the two streams want to swap lanes -- those in the left lane merge into the right lane and vice versa. A small fraction of the vehicles may want to stay in their lanes. 
This  interchange lane swap (ILS) scenario has several interesting aspects from the automated driving point of view:
\begin{itemize}
\item There is no agreed upon priority as most vehicles change lanes.
\item Vehicles use both steering and acceleration to adjust timing for lane changes -- it is a 2-DoF scenario.
\item The lane change is mandatory. It is quite inconvenient if it is not executed within the allocated space (e.g., taking the next highway exit and driving back).
\end{itemize}
One could treat this as a 1-DoF merge and allocate both lanes to a single vehicle, but it would not be speed/energy efficient. In Section \ref{results}, we show that the ILS can be handled without slowdowns even though the negotiation does not commence before the red rectangle zone.

For the ILS task we consider a decentralized Control Barrier Function (CBF) based algorithm. CBFs  have been introduced to assure safety of controlled dynamical system \cite{ames_IEEE17, ames_ECC19}. They have been used for merge control of CAVs, e.g. \cite{liu, xu_feasibility, xiao2024, deshpandeMECC}. For ILS, we use the decentralized Predictor-Corrector CBF algorithm (PCCA) introduced in \cite{santillo}. The algorithm specifically takes into account other agents' known (i.e., third party) constraints. A lack of accounting for third party constraints hampers pure decentralized algorithms without explicit priority framework. We tried the pure decentralized algorithm for ILS and found that, very often, vehicles did not complete the lane change in time and sometimes ran  off the road. With  pure decentralized approaches, CBF constraints have to be soft because feasibility may be an issue (see, for example, \cite{xu_feasibility}).

One important feature of the PCCA CBF-based Safety Filter is that it {\em negotiates} lane swaps with other agents. It has been shown in \cite{jankovicTCST, jankovicAR} that the PCCA algorithm liveness (agility, responsiveness) is related to the inter-agent system instability. Contributions of this paper include showing that an instability-driven approach (IDA) provides excellent performance and robustness in ILS. Moreover, in contrast to 1-DoF cases in \cite{jankovicTCST, jankovicAR}, the speed of instability can be adjusted with a key tuning parameter: the relative cost of steering vs braking in the CBF Quadratic Program (QP). Another contribution is a proof of feasibility for the PCCA QP with elliptical CBFs. This feasibility seems to be one key ingredient in avoiding harsh control reversals. 

To illuminate the role of instability, we consider IDA-fast, with faster instability, and IDA-slow with slower instability. The third algorithm considered (also PCCA-based) relies on virtual guard rails (VGR) to push the vehicles into desired lanes.  
Monte Carlo (MC) simulations compare IDA-fast, IDA-slow, and VGR in the nominal case. Only the IDA-fast is able to perform lane swaps within the 120m highway stretch. One unexpected outcome is that the more agile IDA controllers actually use milder controls, thus improving  passenger comfort and predictability to others. 
In the case when one (randomly selected) vehicle is not reacting to others (the non-responding agent), the IDA-fast still provided very good collision avoidance capability thanks to its instability-induced responsiveness. 

\subsection{Assumptions}
\label{assumptions}
For high-speed, high-density interchange traffic we assume that vehicles are capable of fast acceleration/deceleration and (less critically) steering response. The acceleration limits are set using base-level models of the electric Ford Mach-E and Tesla Model 3 vehicles as benchmarks. 

Situational awareness is provided by the standard vehicle-to-vehicle (V2V) Basic Safety Message (BSM) broadcast by each vehicle at 10Hz frequency as specified in SAE J2735 standard \cite{sae}. The BSM information includes position, velocity, acceleration, heading, steering angle, size, etc. One advantage of V2V is that it is not sensitive to a loss of information due to occlusions. The IDA and VGR do not need a central coordinator. IDA, in particular, works well with the V2V range of 80m (see Subsection \ref{sec:V2V_range}) and is not sensitive to additional signal latency (e.g., BSM at 5Hz).

\section{Control Barrier Functions}
\label{CBF}
In this section we give a brief review of the robust, exponential Control Barrier Functions. The nonlinear system dynamics considered includes disturbance inputs:
\begin{equation} \dot x = f(x) + g(x) u + p(x) w \label{dynamics} 
    \end{equation}
The vector $x\in \R^n$ is the systems state, $u\in {\mathcal U} \subseteq \R^{n_u}$ is the control input and $w\in {\mathcal W} \subseteq \R^{n_w}$ is an external disturbance. 
The system also has an admissible (safe) set ${\mathcal C} = \{ x\in R^n: h(x) \ge 0\}$ with the function $h(x)$ differentiable sufficiently many times. \\
{\bf Definition 1}: The function $h(x)$ is an (exponential) Robust Control Barrier Function (RCBF) if there exists a constant $\lambda > 0$ such that 
\begin{equation}
   \max_{u\in {\mathcal U}}\min_{w\in {\mathcal W}}\{ L_f h + L_g h\ u + L_p h\ w + \lambda h\} >0    \label{ERCBF}
\end{equation}
where the notation $L_f h$ denotes $\frac{\partial h}{\partial x}f$.
The definition combines the concepts of RCBF from
\cite{jankovicRCBF} and exponential CBF from \cite{nguyen}. As long as the control satisfying \ref{ERCBF} is used, $\dot h + \lambda h >0$ and $h(t) > 0, \forall t > t_0$ provided that $x(t_0)$ is in the admissible set $\mathcal C$ --  the set $\mathcal C$ is positively invariant.

In this paper, the control and disturbance inputs  will only appear in the second derivative of the CBF $h(x)$: that is, $L_g h = L_p h = 0$. Following \cite{nguyen}, instead of (\ref{ERCBF}), the requirement for $h$ to be a second-order RCBF is
\begin{equation}
   \barr{l} {\max_{u\in {\mathcal U}}\min_{w\in {\mathcal W}}\{\ddot h(x) + l_1 \dot h + l_0 h \}
=  \\*[2mm]  
   \max_{u\in {\mathcal U}}\min_{w\in {\mathcal W}}
   \{L_f^2 h + L_g L_f h\ u + L_p L_f h\ w + \\*[2mm]
   \hspace*{28mm}  + l_1 L_f h + l_0 h\} >0 }  \label{ERCBF2}
\end{equation}
where $L^2_f h = L_f(L_f h)$, $l_0 = \lambda_1 \lambda_2$ and $l_1 = \lambda_1 +\lambda_2$, with $\lambda_1>0$ and  $\lambda_2>0$, setting  the two roots of the characteristic polynomial $\chi(s) = s^2 + l_1 s + l_0 = 0$ at $\{-\lambda_1, -\lambda_2\}$. In addition, we need to restrict the invariant set to 
${\mathcal C}_\nu = \{ x \in R^n: h(x)>0, \lambda_\nu h(x) \ge -\dot h(x) \}$, where $\nu = 1$ or $\nu = 2$. It is beneficial to use the smaller $\lambda$ (we will assume that it is $\lambda_1$) because it makes the admissible set larger.  

Next we briefly review the design of the safety filter to keep the state in the safe set $\mathcal C_1$. In \cite{jankovicRCBF}, two options for the disturbance $w$ were considered: unknown but bounded and known (estimated). 
In the first case, because the worst case scenario is assumed, the design is conservative. The second case, which is considered here, is less conservative, but the question arises where an estimates of $w$ comes from. This is addressed in the next section. 

The standard approach to employ the CBF to guarantee positive invariance of $\mathcal C$, or $\mathcal C_1$, is to override a baseline, or performance, control $u_0$ using a Quadratic Program (QP). With a disturbance estimate $\hat w \in {\mathcal W}$
assumed available, the QP takes the form
\begin{equation}
\barr{l}{ \min_{u \in {\mathcal{U}}}  \|u - u_0\|^2 \ \ {\rm such \ that} \\*[2mm]
L_f^2 h + L_g L_f  h\ u + L_p L_f  h\ \hat w +  l_1 L_f h + l_0 h = \\*[2mm]
a + b u + b_w \hat w \ge  0  } \label{QP}
\end{equation} 
with $a = L_f^2 h(x)+  l_1 L_f h(x) + l_0 h(x)$, $b = L_gL_f  h(x) $, and $b_w = L_p L_f h(x) $.
By definition of the RCBF, we know that the solution exists, that is, the QP is feasible. 

\section{Vehicle Dynamics and Elliptic CBFs}
\label{e-CBF}
The vehicle dynamics for each agent considered in this paper is the standard simplified bicycle model \cite{rajamani}:
\begin{equation} \barr{l}{\dot x = v \cos\theta \\*[1mm]
    \dot y = v \sin\theta \\*[1mm]
    \dot \theta = \frac{v}{L_w} \delta \\*[1mm]
    \dot v = a_c} \label{bicycle} 
    \end{equation}
where $x, y$ are the (global) cartesian coordinates of the vehicle, $v$ is the vehicle longitudinal velocity, $\theta$ is the orientation relative to the $x$ coordinate, $\delta \approx \tan(\delta)$ is the wheel steering angle, $L_w$ is the wheelbase, and $a_c$ is the vehicle acceleration. It is customary to have the $x$-axis aligned with the road. 

\subsection{Pure-Pursuit baseline controller}
The safety-filter setup requires a baseline controller that usually ignores obstacles and tries to move the vehicle along the desired path. In this paper we employ the standard pure-pursuit algorithm. For the lateral controller, i.e. lane changes, we have used a velocity-based look-ahead  
equal to 1 second travel distance plus 5m. The longitudinal controller just maintains the speed at the desired setting. 

\subsection{Elliptic CBFs}
\begin{figure}
	\centering
	\includegraphics[width=84mm]{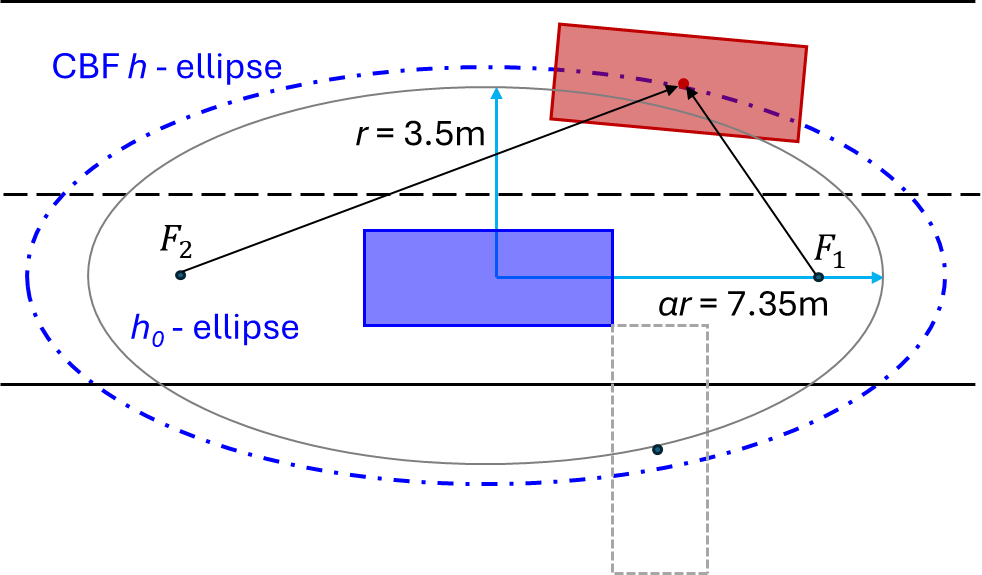}
       \vspace{-3mm}
	\caption{Vehicles and elliptic CBFs with dimensions used in this paper.}
	\label{fig:ellipses}
    \vspace{-3mm}
\end{figure}

CBFs based on ellipse-to-ellipse distance were pursued in  \cite{verginis, funada}. Here, we use an ellipse-to-center approach more amenable to analysis of feasibility and (in)stability. The vehicles are protected by elliptic CBFs so that the center of the other vehicles is outside of the ellipse as depicted in Figure \ref{fig:ellipses}. The size of the solid-line $h_0$-ellipse is chosen such that a collision is prevented under the worst-case (very unlikely) orthogonal orientation as shown. 
The controller uses the CBF dash-dot $h$-ellipse (enlarged to $3.8 \times 8.36$m) with a margin of safety to handle inter-agent  uncertainties and modeling approximations (both discussed below).

 The CBF function $h(x)$ is based on the  definition of the ellipse as the set of points with the sum of distances from the two focal points $F_1$ and $F_2$ equal to the length of the major axis $2\alpha r$ ($r$ is the semi-minor axis and $\alpha$ is the ratio of the major and minor axes). The focal points are at the distance $\rho = r \sqrt{\alpha^2 -1}$ from the ellipse center along the major axis as shown in Fig \ref{fig:ellipses}. If the orientation of the vehicle $i$ is $\theta_i$, the focal points are located at $F_{i1} = X_i +\rho \varphi_i$ and $F_{i2} = X_i -\rho \varphi_i$, with $X_i=[x_i,y_i]^T$ the center of the vehicle and $\varphi_i = [\cos \theta_i, \ \sin \theta_i]^T$. Thus, the elliptical CBF for vehicles $i$ and $j$ is given by:
\begin{equation}
    h_{ij} = \|F_{i1} -X_j\| + \|F_{i2} -X_j\| -2\alpha r \label{h_ellipse}
\end{equation}
With the vehicle dynamics (\ref{bicycle}), and using $\xi_{ij,k} = F_{ik} -X_j, \ k = 1,2$ ($k$ corresponds to the focal point), we obtain
\begin{equation} 
 \dot h_{ij} = \sum_{k = 1}^2 \frac{\xi_{ij,k}^T}{\|\xi_{ij,k}\|}(v_i\varphi_i- v_j\varphi_j)  \label{h_dot}
\end{equation}
To avoid mismatched relative degrees between steering and acceleration inputs, we have assumed that the dynamics of the focal points is the same as that of the center of the vehicle. This approximation neglected terms  proportional to the steering angle $\delta$ that remained small ($\pm 0.02$rad) in the ILS scenario considered.
The barrier margin has been able to handle the approximation as confirmed in Section \ref{results}. 

Now, both control inputs appear in the second derivative of the CBF: 
\begin{equation}
\barr{l}{\ddot h_{ij} = \sum_{k = 1}^2 \frac{1-\cos^2\beta_{ij,k}}{\|\xi_{ij,k}\|} \|(v_i\varphi_i- v_j\varphi_j)\|^2  + \\*[2mm]
\hspace{7mm}+\sum_{k = 1}^2 \frac{\xi_{ij,k}^T}{\|\xi_{ij,k}\|} (\gamma_{i} u_i+ \gamma_j u_j) } \label{hddot}
\end{equation}
where $\beta_{ij,k}$ is the angle between the vectors $\xi_{ij,k}, \ k =1,2$ and $v_i\varphi_i- v_j\varphi_j$, while $\gamma_i = [v_i^2/L_w \varphi'_i; \ \varphi_i]$, $\varphi'_i=[-\sin\theta, \cos \theta]^T$, $u_i = [\delta_i, a_{ci}]^T$ and the same for $\gamma_j, \ u_j$. 


\subsection{Road-boundary CBF}
\label{rb-CBF}
The CBF for the road boundary is very simple. We require that the lateral position $y$ of each agent stays within the road boundaries on the left and right: $y_i \in[rb_r, rb_l]$. Thus, the pair of road  CBF constraints for the agent $i$ are 
\begin{equation}
    h_{ir} = y-rb_r \ \ {\rm and} \ \ h_{il} =- y + rb_l \label{h_road}
\end{equation}
The CBF constraint is obtained by combining $h_{ir} (h_{il})$
and its first two derivatives to obtain
\[ a_{ir} + b_{ir} u_i \ge 0 \ \ \ (a_{il} + b_{il} u_i \ge 0)\] 
where $a_{ir} = l_1 v_i \sin\theta_i + l_0 h_{ir}, \ \  b_{ir} = [v_i^2 \cos\theta _i /L_w,\ \sin \theta_i]$ and similarly for the left road boundary.

The main difference (besides tuning) between the IDAs and the VGR is that lane-changing agents have a road boundary $rb = d_0 + d_1 \tan^{-1}(d_3(x-d_4))$ forcing them to change lanes (the constraint is soft, though). The parameters $d_i$ are tuned to produce guardrails shown in Fig. \ref{fig:lane_swap}.

\section{CBF SAFETY  FILTERS -- PCCA}
\label{SF}
Various versions of the CBF (and the related Potential Field) based methods have been considered in the highway merge literature. The most popular is the ``pure" decentralized algorithm in which each agent computes its own action and takes full responsibility for  avoiding other agents (see \cite{ hayashi, xiao2020decentralized, xiao2024}). The main advantage is computational simplicity, while disadvantages for the ILS problem were discussed in the Introduction.

Centralized CBF algorithms take into account everyone's constraints and have everyone's control action at their disposal. One such algorithm has been  considered in \cite{deshpandeMECC} for the 1DoF (i.e. on-rail) merge scenario. However, moving from 1DoF merge to 2DoF ILS would require expanding V2V (or vehicle-to-infrastructure) message content well beyond BSM. 

In this paper we use the third option: decentralized Predictor-Corrector CBF Algorithm (PCCA), see \cite{santillo, jankovicTCST}.
The algorithm is actually quasi-centralized, that is, each agent independently computes everyone's control action based on the information it has (e.g., BSM does not provide desired speed and lane change intent). Each agent implements its own computed action while the others are local copies. Unavoidable disagreements are reconciled with predictor-corrector loops within the RCBF setting: differences between the observed (BSM communicated) actions of other agents and the corresponding local copies are treated as known disturbances $w_{ij}$. The first index (e.g., ``$i$'') corresponds to the agent computing them. Following Section \ref{CBF}, the PCCA Safety Filter for agent $i$ is given by 
\begin{equation}
  \barr{l}{ \min_{u \in {\mathcal{U}}}  \|u_{ii} - u_{i0}\|^2_{S_i} + \sum_{j=1, j\not = i}^{N_a}  \|u_{ij} \|^2_{S_j}\ \ {\rm such \ that} \\*[2mm]
a_{jk} + b_{jk} (u_{ij} + w_{ij}) + b_{kj} (u_{ik} + w_{ik}) \ge  0 , j\not = k \\*[2mm] 
a_{kq} + b_{kq} ( u_{ik} + w_{ik} ) \ge  0, \ q =\{``r", ``l"\} }  \label{PCCA}
\end{equation}
where  $\|\psi\|^2_{S_j} = \psi^T S_j \psi$, while $S_j= {\rm diag}\{1, s_{a j}(v_j) \}$, our instability tuning knob,  sets relative sensitivity between steering and acceleration. The set ${\mathcal U}=\{u \in {\R}^{2N_a}: \underbar {u} \le u_i \le \bar u \} $ with $\underbar u$ and $\bar u$ vectors of the upper and lower limits on the control variables. In the implementation, different limits are used for the ego controls and for the local copies of other agent's controls (see Remark \ref{remark1} below). 

The solution to the QP (\ref{PCCA}) produces the control action $u_i^* \in {\R}^{2N_a}$ that agent-$i$ believes everyone should apply. It then goes ahead and implements its own computed action $ u^*_{ii}$. The "known" disturbances -- differences between the local copies and the control action other agents actually applied  -- are filtered to break the algebraic loop:
\begin{equation}
    \dot w_{ij} = \frac{1}{\tau} (-w_{ij}+u^*_{jj} - u^*_{ij}) \label{wij}, \ \ j = 1,\ldots, N_a
\end{equation} 

\noindent
\begin{proposition} \label{feas} 
Without the $\mathcal U$ (control limit) constraints, the QP for agent $i$ defined by (\ref{PCCA}) is always feasible and one feasible action is $u_{ij}^f = [0, -\lambda_1  v_j]^T - w_{ij}$ for all $j \in \{1,\ldots, N_a\}$.
\end{proposition}

{\bf Proof:} \ Inside the admissible set for the 
second order CBF $\mathcal C_1$, $\lambda_1 h \le -\dot h$ (see Section \ref{CBF}), which implies that $l_1 \dot h + l_0 h \ge \lambda_1 \dot h$. The proposed feasible action $u_{ij}^f = [0, -\lambda_1  v_j]^T - w_{ij}$ cancels $w$'s as well as $\lambda_1 \dot h$ term and we have $a_{jk} + b_{jk}(u_{ij} + w_{ij})
+ b_{kj}(u_{ik} + w_{ik}) \ge L_f^2 h \ge 0$ $\forall j,k$, as computed by agent $i$.  The same consideration applies to 
the line constraints except that $L_f^2 h \equiv 0$. \hfill $\nabla$
\ \\
\noindent
\begin{remark} \label{remark1} The algorithm (\ref{PCCA}) retains enough flexibility to handle control  $\mathcal U$ constraints as well. The feasible action consists of 0 steering and braking proportional to agent's speed, after the disturbance terms are canceled. The disturbance term for the ego agent $w_{ii}$ is always $0$: its own computed action is the one applied. Thus, it only needs to have sufficient control authority to brake proportionally to its speed. Non-ego agent controls, in addition, need to cancel corresponding $w$'s. We accommodated this  by selecting wider box constraint limits ($\underbar {\it u}$ and $\bar u$) for the non-ego agent controls (we used $1.8\times$ wider box). 
Because they are local copies, the impact is indirect.  

 We have been able to run IDA (both slow and fast) in extensive MC simulations with no slack variables for the QP and with no infeasibility observed. For  VGR, the feasibility proof does not work because $L^2_f h \not \ge 0$. Indeed, we had to use soft constraints in MC simulations to prevent infeasibility. 
\end{remark}

\subsection{Instability}
\noindent
In 1-DoF analysis \cite{jankovicTCST, jankovicAR}, the PCCA algorithm has been shown to produce "gridlock" equilibria that are {\em unstable}. The 2-DoF case considered here turned out to be different in some key aspects. To illustrate the situation, we consider a particular trajectory in the 2-agent scenario where the agents run in parallel at the same constant desired speed $v_0$ while attempting to swap lanes. We assume that the baseline steering $\pm \delta_0$ commands are constant (positive for agent 1 in the right and negative for agent 2 in the left lane), while the acceleration tries to keep agent's speed constant $a_{c,i0} = -\kappa(v_i - v_0)$. The solution to the PCCA QP (\ref{PCCA}) is available in closed form. Due to space limitations, only an outline is provided -- the system is of high order (8 for the agents' states and 4 for the $w$ terms) while the derivation is tedious but relatively straightforward. 

Linearizing the system around the constant speed trajectory, the state splits into 4 parts: two lateral and two longitudinal. Assuming that the $w$-dynamics is very fast ($\tau $ close to 0), we used Singular Perturbation method to remove these states from consideration, the  remaining states split into four groups:  $ z_{lg}^\Delta = [x_1 - x_2, v_1 - v_2]$, $z_{lg}^\Sigma= [x_1 + x_2, v_1 + v_2]$, $z_{lat}^\Delta  = [y_1 - y_2, \theta_1 - \theta_2]$, and $z_{lat}^\Sigma  = [y_1 + y_2, \theta_1 + \theta_2]$. In the transformed state space, linearized around the moving equilibrium, the  $ z_{lat}^\Delta$  and $z_{lg}^\Sigma$  states are decoupled from the rest and each other. The dynamics of the former are that of the CBF $h$ with $\{ -\lambda_1, -\lambda_2\}$ as the eigenvalues of the state matrix. The later has eigenvalues at $\{0, -\kappa\}$. Both subsystems are stable. The $ z_{lat}^\Sigma$  and $z_{lg}^\Delta$ subsystems remain coupled and the four eigenvalues of the joint system state matrix are 
\[ \left\{0, 0, -\frac{\kappa}{2} \pm \sqrt{\frac{\kappa^2}{4}+4\frac{2\delta_0}{s_a r v_0^2}(\frac{\delta_0 v_0}{L_w} + \frac{L_w}{\alpha^2})} \right\} \]
 where, recall, $s_a= s_a(v_0)$ is the sensitivity tuning parameter that weighs steering vs acceleration action for collision avoidance.  
Note that one eigenvalue is positive regardless of the controller tuning and, thus, the $ (z_{lat}^\Sigma, \ z_{lg}^\Delta)$ dynamics is unstable. 
Prominent features include:
\begin{enumerate}
    \item The instability depends on the baseline speed control gain. A strong speed control feedback results in slow divergence and inability of the vehicles to complete the lane swaps in time. 
    \item In contrast to PCCA (and Centralized) controller in the holonomic agent case \cite{jankovicTCST, jankovicAR}, here we have a tuning parameter $s_a$. Because of the $v_0^2$ in the denominator of the instability defining term, we set $s_a(v) = 1/(c_0 + c_2 v^2 +c_3 v^3) $: $c_0$ is a small offset to preserve strict convexity of the QP, while the cubic term is added for faster instability the faster the vehicles travel. In  Section \ref{results}, the IDA-fast tuning of $s_a(v)$ sets the unstable eigenvalue at  $2.6 s^{-1}$ at $10$mph; $3.1 s^{-1}$ at $20$mph, and $3.5 s^{-1}$ at $30$mph (with $\delta_0 = 0.015$ rad). The IDA-slow tuning sets the eigenvalues at 1/2 of the IDA-fast (this is their only difference).   
\end{enumerate}

\section{EVALUATION AND RESULTS}
\label{results}
In this section we consider homogeneous traffic (vehicles of the same size $L_v \times W_v = 4.7 \times 1.85$m  and capabilities, e.g., Ford Mach-E or Tesla Model 3) entering the ILS segment. The acceleration limits are $[-8, 4]m/s^2$,  within capabilities of the two models. The steering is limited to $\pm \pi/7$ rad, but never gets anywhere close. The lane width is $3.5$m.  

For the nominal  MC runs, the two IDA versions could have used hard CBF constraints without being infeasible. It is actually safer to use  "soft" constraints that are needed in the non-responding agent case. VGR needs soft constraints even in the nominal case. Thus, all versions used soft (but fairly stiff) constraint with the slack-variables weight equal to  20,000 for the inter-agent CBFs and 1,000 for the lane boundaries. The baseline controller is the standard pure pursuit as discussed in Section \ref{CBF} with $\kappa = 0.7s^{-1}$. The $h_{ij}$ CBF constraint $\lambda$'s are $\{0.4, 4\}$.

\begin{figure*}
	\centering
    \includegraphics[width=170mm]{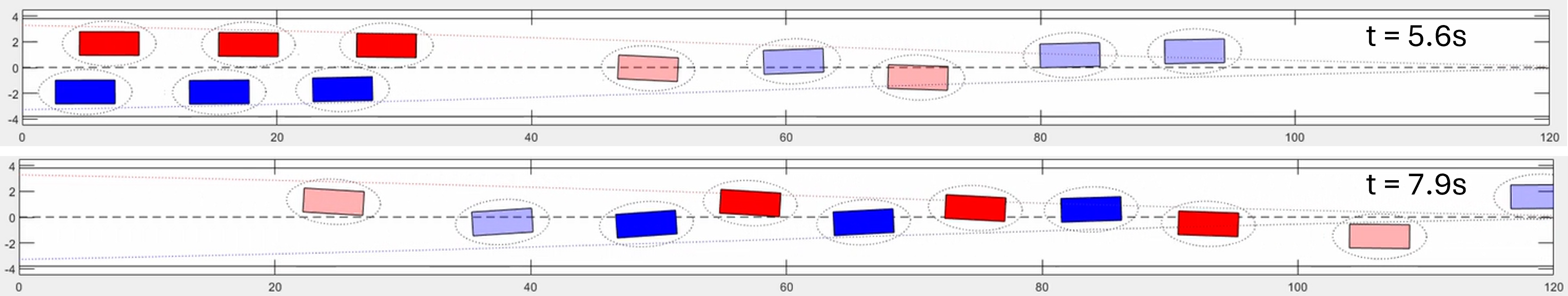}
    \vspace{-1mm}
	\caption{IDA-fast interchange lane swap: the non-transparent blue and red rectangles correspond to the same group of 6 vehicles at two time instants 2.3s apart. The "guardrail" lines are shown, but are not active for the IDA.  }
	\label{fig:lane_swap}
\end{figure*}

The fast instability of IDA-fast provides the group of vehicles a superior level of responsiveness. 
While this is easier to appreciate in a video, we illustrate the capability by a contested, high-speed, high-density lane swap in 120m highway stretch that approximately corresponds to the dashed rectangle in Fig. \ref{fig:interchange}. The traffic situation depicted in Fig. \ref{fig:lane_swap} focuses on a group of 6 vehicles  that arrive at the interchange section running almost in parallel. In this case, for clarity, all the vehicles are changing lanes.  The ellipses shown Fig. \ref{fig:lane_swap} are just for illustration. They are 1/2 size of the $h_0$ ellipse shown in Fig \ref{fig:ellipses}.

\begin{figure} 
	\centering
    \includegraphics[width=\columnwidth]{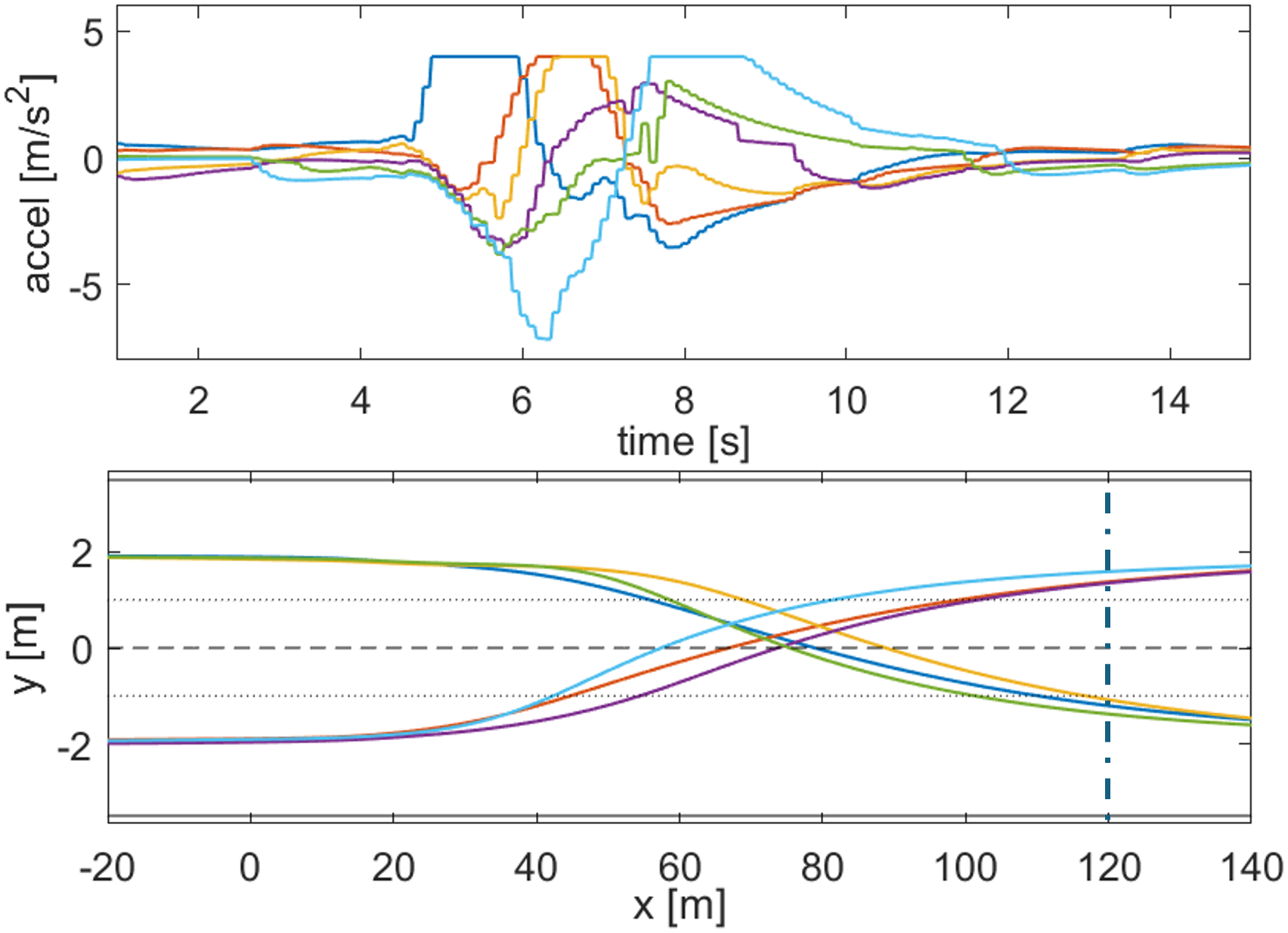}
    \vspace{-5mm}
	\caption{IDA-fast run. Top plot: acceleration profiles of the 6 vehicles in focus. Bottom plot: their velocities in the $(x,y)$ plane.  }
	\label{fig:acell}
\end{figure}

 The average speed of the group at $t = 5.6$s is 55.2mph. Within a very short period of time they manage to position themselves for the lane swap.  The average speed at $t = 7.9$s is 54.9mph, very close to their speed at the entrance to the interchange zone. Their accelerations are shown in the top plot in Fig \ref{fig:acell}. The bottom plot shows the trajectories of the 6 vehicles in the $(x,y)$ plane and the completion of the lane changes just in time. 

Here are a few key points about the IDA:
\begin{enumerate}
    \item Despite lack of assigned priority or means of coordinating the motion, we note well synchronized response of the vehicles. They are all required to avoid colliding with others and are all aware of third-party constraints. As a result, there  are no visible signs of string instability where each following vehicle slows down a little later a little more and  than the one ahead of it. 
    \item  The vehicles don't know that others are changing lanes. Each signals their intent through their motion and actions (broadcast by BSM) while PCCA $w$-loops quickly reconcile differences, keeping the minimum value of $h_{ij}^0$ over all agent pairs and all times above $0.24$m.
    \item While the controller has to be aggressive to accomplish the lane swap, there is no sudden change of mind in acceleration or steering. The maximal accleration change in one sample interval ($100$ms) is $\max \Delta a_{c} = 2.35 m/s^2$ and there are only 4 such events that exceed $2m/s^2$ providing some predictability of agents' actions based on BSM updates at $100$ms.
    \item  The vehicles complete this contested lane swap within the allocated space.  All the agents stay inside their desired lanes (outside the dotted lines placed at 1/2 vehicle width from the center line) before the vertical ``finish line" at $120$m, bottom plot in Fig. \ref{fig:acell}. 
\end{enumerate}

\subsection{Virtual Guard Rails}
\label{sec:VGA}
As mentioned above, one intuitive approach to handle mandatory lane changes is to include virtual guardrails (VGR) depicted in Fig. \ref{fig:lane_swap} (inactive for IDA). The guardrails provide CBF constraints on the left or right as appropriate. Beside the guardrails, the only difference from the IDA controller is that the tuning of the sensitivity factor $s_a$  produces a much milder unstable eigenvalue at $0.13 s^{-1}$ at 20mph (we retained this original tuning of the VGR controller that was developed before IDAs). 

\begin{figure} 
	\centering
    \includegraphics[width=\columnwidth]{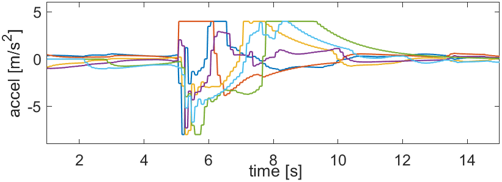}
    \vspace{-5mm}
	\caption{Virtual Guard Rails acceleration profiles for the same 6 vehicles and the same initial conditions used in Fig. \ref{fig:acell}  }
	\label{fig:VGR}
     \vspace{-5mm}
\end{figure}

Fig. \ref{fig:VGR} shows the acceleration profile for the VGR approach that can be compared to the IDA-fast in the top plot in Fig. \ref{fig:acell}. The VGR accelerations are much harsher with the maximum acceleration delta between two consecutive samples equal to $9.2m/s^2$. The reason is not difficult to explain: relying on guardrails gets vehicles pushed into the guardrail "funnel" that causes sudden onset of actions and reactions.  

\subsection{Monte Carlo Simulations}
\label{sec:MonteCarlo}
 We ran 100 Monte Carlo simulations with IDA-fast, IDA-slow, and VGR for 16 vehicles with (identical) random initial positions and initial/desired speeds in the range 20 to 25m/s providing the average traffic density of 3500 veh/h per lane (about 1 vehicle every second). This MC set is a randomized version of the run shown above except that $15$\%   of the vehicles, on average, are randomly selected to go straight and not change lanes. Only the ego vehicle knows if it is changing lanes or not. 


\begin{table}
	\caption{Comparison metric for two IDAs and VGR in 100 MC runs}	
    \vspace{-0.1in}
	\begin{tabular}{|l|cc|c|c|c|}
	\cline{1-6}
	                                & \multicolumn{2}{c|}{Average}             &         max       &      &  Average        \\
	                 Method   & speed                                   & brake loss & OOB & max $\Delta a_c$ &  $\Delta a_c > 2$  \\ 
                      & $[m/s]$                                  & $[Wh/km]$  & $[m]$  & $[m/s^2]$  & \# \\ \hline
	\multicolumn{1}{|l|}{IDA-fast}      & \multicolumn{1}{c|}{50.4} & \multicolumn{1}{c|}{62}   & N/A    & 5.6        & 11          \\ \hline
	\multicolumn{1}{|l|}{VGR}           & \multicolumn{1}{c|}{50.2} & \multicolumn{1}{c|}{48}   & {\bf 0.5}      & 12          & 39        \\ \hline
    \multicolumn{1}{|l|}{IDA-slow}           & \multicolumn{1}{c|}{50.4} & \multicolumn{1}{c|}{49}   & N/A      & 3.3          & 1        \\ \hline
	\end{tabular}
	\vspace{-0mm}
	\label{tab:nominal}
\end{table}

\begin{figure} [h] 
    \hspace{-6.0mm}
    \includegraphics[width=100mm]{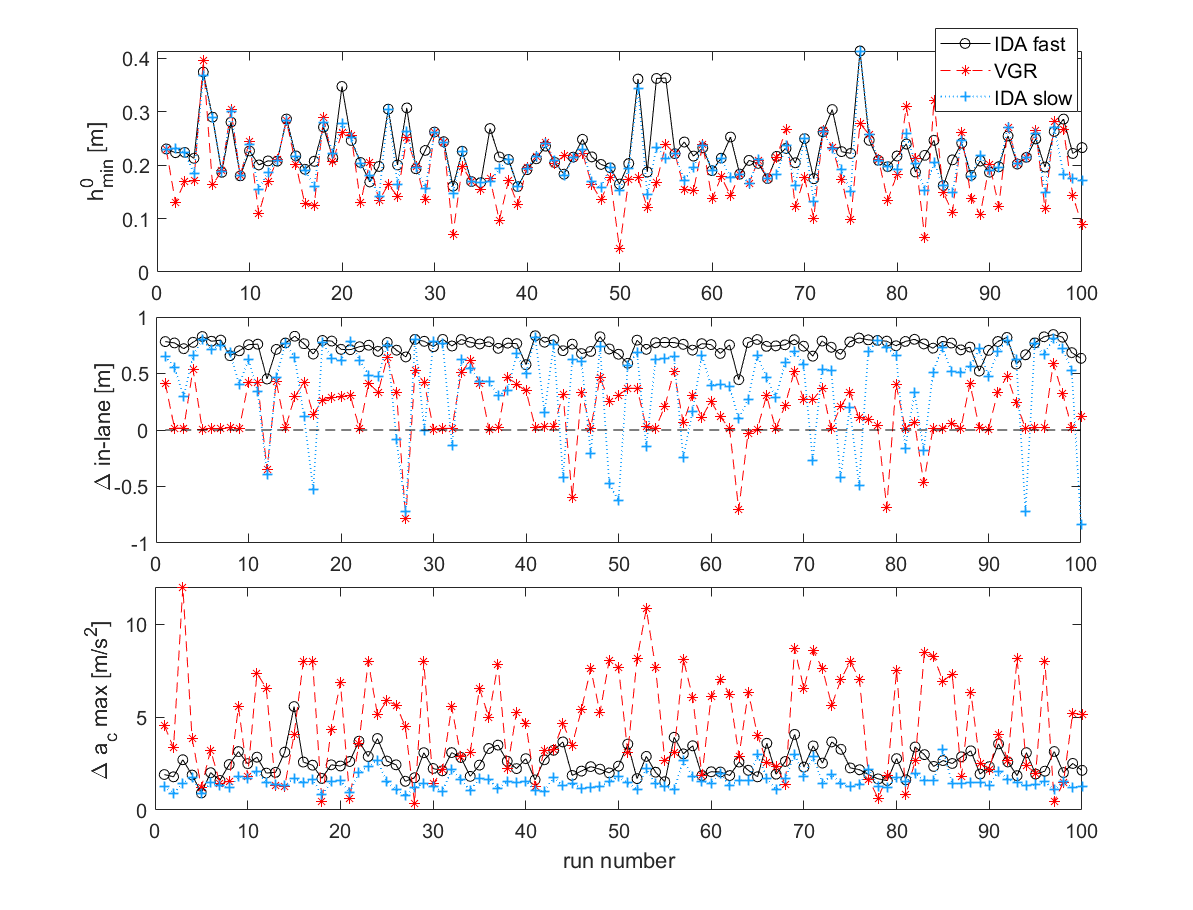}
 \vspace*{-9mm}
	\caption{Top plot: minimal value of all agent-to-agent CBFs over a run; Middle plot: vehicle distance from the assigned-lane boundary at 120m; Bottom plot: the maximal delta-acceleration over one sample period, over all agents.}
	\label{fig:nominal}
    \vspace*{-4mm}
\end{figure}

From Table \ref{tab:nominal}, one can conclude that the IDAs and VGR have comparable speed and energy efficiency. The initial and desired average speed is 50.6mph, so the vehicle group  decelerates very little on average. The "OOB" is out-of-bounds metric meaning that, in 100 MC runs, one VGR controlled vehicle went out of the road by 0.5m (bold font because it is highly undesirable). The top plot in Fig. \ref{fig:nominal}  shows controllers maintaining the $h_0$ safety, though the IDAs are a little better at it than the VGR.  The last two items in the table and the bottom plot in Fig. \ref{fig:nominal}  show that IDAs have lower delta-acceleration (jerk) and should be more predictable and more comfortable. We note in particular the mild behavior of the IDA-slow. VGR delta-acceleration goes up to $12m/s^2$, which is a complete reversal between max acceleration and max braking. VGR and IDA-slow sometimes do not complete the lane change in time (Fig. \ref{fig:nominal}, middle plot). They both need more time/space. Finally, in this set of 100 simulation episodes, the compute (loop) time averages 5.2ms on a laptop (i7-3800H, 2.5GHz) running Matlab quadprog solver. The longest loop time observed was 23ms for VGR, 13.6ms for IDA-slow, and 12.6ms for IDA-fast.  
 
\subsection{Non-responding agent}
\label{sec:NRA}
To address robustness, we consider an agent that ignores all others and performs its lane-change maneuver as if it is alone on the road.  For each of the 30 MC runs, the non-responding agent (NRA) is randomly selected from the pool of 16 vehicles with all else being equal to the nominal MC set. Other agents are not aware there is a not-responding agent and are not trying to detect aberrant behavior. 

\begin{figure} 
    \hspace{-6mm]}
    \includegraphics[width=95mm]{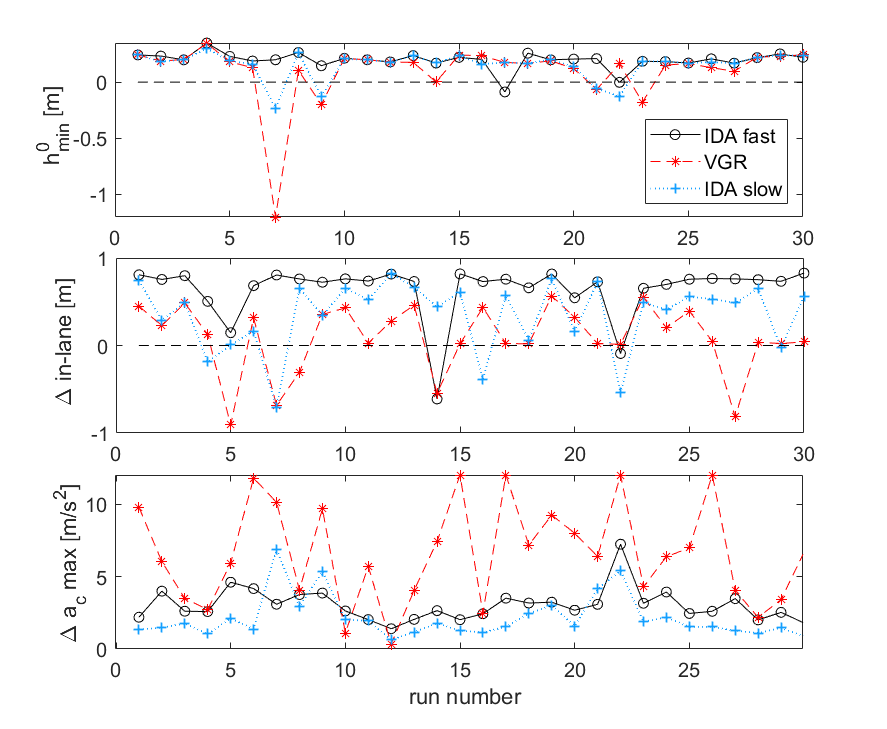}
  \vspace*{-8mm}
	\caption{Non-responding agent case. Top plot: minimal value of all the agent-to-agent CBFs over a run; Middle plot: vehicle distance from the desired lane boundary at 120m; Bottom plot: the maximal delta-acceleration over one sample period, over all agents and times. }
	\label{fig:NRA}
\end{figure}

The  agility of IDA continues to provides noticeable benefit with minimal CBF violations (the top plot in Fig. \ref{fig:NRA}). IDA-fast has only one incomplete merge  (middle plot in Fig. \ref{fig:NRA}).  The delta acceleration remains well controlled. 

\subsection{V2V range and update-rate impact}
\label{sec:V2V_range}
In congested traffic, V2V communication may be impacted by range limits and latency. We reran the MC simulations looking for effects of limiting the V2V range to 50m or 80m and reducing sampling rate to 5Hz (0.2s). The results for IDA-fast and VGR are shown in Table \ref{tab:V2V_effects}.  At $50$m V2V range, there are a few incomplete lane swaps ("incomplete LS") for IDA-fast. Otherwise, at $80m$ range and at $0.2s$ IDA-fast performs well. The increased $\Delta a_c$  is due to the change from $0.1s$ to $0.2s$ update. On the other hand, VGR has problems with reduced range and, especially, slower update rate ($h_0 = -1.3$m would produce a collision).

\begin{table}
	\caption{100 MC runs exploring effect of V2V range and latency}	
    \vspace{-0.1in}
	\begin{tabular}{|l|c|c|c|c|c|}
	\cline{1-6}
	                 Method  and                                   & min $h_0$ & incomp- & OOB & max $\Delta a_c$\! &  $\Delta a_c > 2$\!  \\ 
                    use case  &   $[m]$      &          lete  LS    \#  & $[m]$  & $[m/s^2]$  & \#  \\ \hline
	\multicolumn{1}{|l|}{IDA-fast\  50m}       & \multicolumn{1}{c|}{0.15}  & {\bf 3} & N/A    & 4.7        & 11          \\ \hline
	\multicolumn{1}{|l|}{VGR \ \ \  \ 50m}            & \multicolumn{1}{c|}{\bf --0.07}  & {\bf 5}  & {\bf 0.4}      & 12          & 41        \\ \hline
    \multicolumn{1}{|l|}{IDA-fast    80m}          & \multicolumn{1}{c|}{0.14} & 0  & N/A      & 5.9          & 12        \\ \hline
    \multicolumn{1}{|l|}{VGR \ \  \ \  80m}      & \multicolumn{1}{c|}{0.01} & {\bf 5}  & {\bf 0.5}   & 12        & 39          \\ \hline
	\multicolumn{1}{|l|}{IDA-fast 0.2s}          & \multicolumn{1}{c|}{0.07} & 0  & N/A       &   9.6      & 90        \\ \hline
    \multicolumn{1}{|l|}{VGR \ \ \ \ 0.2s}          & \multicolumn{1}{c|}{\bf --1.3} &  {\bf 4} &  {\bf 0.9 }    & 12         & 102        \\ \hline
	\end{tabular}
	\vspace{-3mm}
	\label{tab:V2V_effects}
\end{table}

\section{Conclusion}
This paper showcases capabilities of recently developed CBF algorithms to perform high speed, high density interchange lane swaps with decentralized controllers. The key role of instability is analyzed and the resulting agility, safety, and robustness illustrated by simulations. The conclusion is that instability-driven approach performs better, is more agile and more robust than forcing the lane change. The problem with the latter is that feasibility cannot be a-priori guaranteed which in turn causes occasional harsh control reversals. 


\section{Acknowledgment}

The authors would like to thank Stas Gankov, Mike Brown, Scott Hotz, JoLyn Swain, and Piyush Bhagdikar, all from Southwest Research Institute for contributing to research that led to this paper.


\bibliographystyle{IEEEtran}
\bibliography{IEEEabrv,mybibfile}

\begin{thebibliography}{10}
\providecommand{\url}[1]{#1}
\csname url@rmstyle\endcsname
\providecommand{\newblock}{\relax}
\providecommand{\bibinfo}[2]{#2}
\providecommand\BIBentrySTDinterwordspacing{\spaceskip=0pt\relax}
\providecommand\BIBentryALTinterwordstretchfactor{4}
\providecommand\BIBentryALTinterwordspacing{\spaceskip=\fontdimen2\font plus
\BIBentryALTinterwordstretchfactor\fontdimen3\font minus \fontdimen4\font\relax}
\providecommand\BIBforeignlanguage[2]{{%
\expandafter\ifx\csname l@#1\endcsname\relax
\typeout{** WARNING: IEEEtran.bst: No hyphenation pattern has been}%
\typeout{** loaded for the language `#1'. Using the pattern for}%
\typeout{** the default language instead.}%
\else
\language=\csname l@#1\endcsname
\fi
#2}}

\bibitem{scarinci}
R.~Scarinci and B.~Heydecker, ``Control concepts for facilitating motorway on-ramp merging using intelligent vehicles,'' \emph{Transport reviews}, vol.~34, no.~6, pp. 775--797, 2014.

\bibitem{rios-torres}
J.~Rios-Torres and A.~A. Malikopoulos, ``A survey on the coordination of connected and automated vehicles at intersections and merging at highway on-ramps,'' in \emph{IEEE Transactions on Intelligent Transportation Systems}.\hskip 1em plus 0.5em minus 0.4em\relax IEEE, 2017, pp. 1066--1077.

\bibitem{mccartt}
A.~T. McCartt, V.~S. Northrup, and R.~A. Retting, ``Types and characteristics of ramp-related motor vehicle crashes on urban interstate roadways in northern virginia,'' \emph{Journal of Safety Research}, vol.~35, no.~1, pp. 107--114, 2004.

\bibitem{mukai}
M.~Mukai, H.~Natori, and M.~Fujita, ``Model predictive control with a mixed integer programming for merging path generation on motor way,'' in \emph{2017 IEEE Conference on Control Technology and Applications (CCTA)}.\hskip 1em plus 0.5em minus 0.4em\relax IEEE, 2017, pp. 2214--2219.

\bibitem{xu}
H.~Xu, S.~Feng, Y.~Zhang, and L.~Li, ``A grouping-based cooperative driving strategy for cavs merging problems,'' \emph{IEEE Transactions on Vehicular Technology}, vol.~68, no.~6, pp. 6125--6136, 2019.

\bibitem{hayashi}
Y.~Hayashi and T.~Namerikawa, ``Merging control for automated vehicles using decentralized model predictive control,'' \emph{IEEE Transactions on Control Systems Technology}, pp. 268--273, 2018.

\bibitem{rios-torres_merge}
J.~Rios-Torres and A.~A. Malikopoulos, ``Automated and cooperative vehicle merging at highway on-ramps,'' \emph{IEEE Transactions on Intelligent Transportation Systems}, vol.~18, no.~4, pp. 780--789, 2017.

\bibitem{xiao2020decentralized}
W.~Xiao and C.~G. Cassandras, ``Decentralized optimal merging control for connected and automated vehicles with optimal dynamic resequencing,'' in \emph{2020 American Control Conference (ACC)}.\hskip 1em plus 0.5em minus 0.4em\relax IEEE, 2020, pp. 4090--4095.

\bibitem{xu_feasibility}
K.~Xu, W.~Xiao, and C.~G. Cassandras, ``Feasibility guaranteed traffic merging control using control barrier functions,'' in \emph{2022 American Control Conference (ACC)}, 2022, pp. 2309--2314.

\bibitem{deshpandeMECC}
S.~Rajakumar~Deshpande and M.~Jankovic, ``Energy-efficient merging of connected and automated vehicles using control barrier functions,'' \emph{IEEE Transactions on Control Systems Technology}, 2025.

\bibitem{ames_IEEE17}
A.~D. Ames, X.~Xu, J.~W. Grizzle, and P.~Tabuada, ``Control barrier function based quadratic programs for safety critical systems,'' \emph{IEEE Transactions on Automatic Control}, vol.~62, no.~8, pp. 3861--3876, 2017.

\bibitem{ames_ECC19}
A.~D. Ames, S.~Coogan, M.~Egerstedt, G.~Notomista, K.~Sreenath, and P.~Tabuada, ``Control barrier functions: Theory and applications,'' in \emph{18th European Control Conference}, 2019, pp. 3420--3431.

\bibitem{liu}
H.~Liu, W.~Zhuang, G.~Yin, and et~al, ``Decentralzied on-ramp merging control of connecetd and autmated vehicels in the mixed traffic using control barrier functions,'' in \emph{IEEE Intelligent Transportation Systems Conf.}\hskip 1em plus 0.5em minus 0.4em\relax IEEE, 2021, pp. 1125--1131.

\bibitem{xiao2024}
W.~Xiao and C.~G. Cassandras, ``Decentralized optimal merging control for mixed traffic with vehicle inference,'' in \emph{2024 American Control Conference (ACC)}.\hskip 1em plus 0.5em minus 0.4em\relax IEEE, 2024, pp. 1468--1473.

\bibitem{santillo}
M.~Santillo and M.~Jankovic, ``Collision free navigation with interacting, non-communicating obstacles,'' in \emph{2021 American Control Conference (ACC)}, 2021, pp. 1637--1643.

\bibitem{jankovicTCST}
M.~Jankovic, M.~Santillo, and Y.~Wang, ``Multiagent systems with cbf-based controllers: Collision avoidance and liveness from instability,'' \emph{IEEE Transactions on Control Systems Technology}, vol.~32, no.~2, pp. 705--712, 2023.

\bibitem{jankovicAR}
M.~Jankovic, ``Instinctive negotiation by autonomous agents in dense, unstructured traffic: A controls perspective,'' \emph{Annual Review of Control, Robotics, and Autonomous Systems}, vol.~7, 2024.

\bibitem{sae}
{SAE International Technical Standard}, ``{V2X} communications message set dictionary,'' Rev. 2024, sAE Standard J2735.

\bibitem{jankovicRCBF}
M.~Jankovic, ``Robust control barrier functions for constrained stabilization of nonlinear systems,'' \emph{Automatica}, vol.~96, pp. 359--367, 2018.

\bibitem{nguyen}
Q.~Nguyen and K.~Sreenath, ``Exponential control barrier functions for enforcing high relative-degree safety-critical constraints,'' in \emph{2016 American Control Conference}.\hskip 1em plus 0.5em minus 0.4em\relax IEEE, 2016, pp. 322--328.

\bibitem{rajamani}
R.~Rajamani, \emph{Vehicle Dynamics and Control}.\hskip 1em plus 0.5em minus 0.4em\relax Springer, NY, 2012.

\bibitem{verginis}
C.~K. Verginis and D.~V. Dimarogonas, ``Closed-form barrier functions for multi-agent ellipsoidal systems with uncertain lagrangian dynamics,'' \emph{IEEE Control Systems Letters}, vol.~3, no.~3, pp. 727--732, 2019.

\bibitem{funada}
R.~Funada, K.~Nishimoto, T.~Ibuki, and M.~Sampei, ``Collision avoidance for ellipsoidal rigid bodies with control barrier functions designed from rotating supporting hyperplanes,'' \emph{IEEE Transactions on Control Systems Technology}, vol.~33, no.~1, pp. 148--164, 2025.

\end{thebibliography}

\end{document}